\documentclass{webofc}
\usepackage[varg]{txfonts}   
\usepackage{bm}
\usepackage{hyperref}
\usepackage{amsmath,epsfig}
\usepackage{subfigure}
\usepackage{amssymb,amsfonts}
\usepackage{latexsym}
\usepackage{epsfig}
\usepackage{amssymb}
\usepackage{graphicx}
 \usepackage{bm}

\usepackage{color}
\usepackage{textcomp}
\usepackage{wasysym}
\def \be {\begin{equation}}
\def \ee {\end{equation}}
\def \ba {\begin{array}}
\def \ea {\end{array}}
\def \bea{\begin{eqnarray}}
\def \eea{\end{eqnarray}}
\begin{document}
\title{QCD critical end point from a realistic PNJL model}
\author{\firstname{Kun} \lastname{Xu}\inst{1,2}\fnsep\thanks{\email{xukun@ihep.ac.cn}} \and
           \firstname{Zhibin} \lastname{Li}\inst{1,2}\fnsep\thanks{\email{lizb@ihep.ac.cn}} \and
           \firstname{Mei} \lastname{Huang}\inst{1,2}\fnsep\thanks{\email{huangm@ihep.ac.cn}}
}

\institute{
Institute of High Energy Physics, Chinese Academy of Sciences, Beijing 100049, China
\and
University of Chinese Academy of Sciences, Beijing 100049, China
}

\abstract{%
With parameters fixed by critical temperature and equation of state at zero baryon chemical potential, a realistic Polyakov--Nambu--Jona-Lasinio (rPNJL) model predicts a critical end point of chiral phase transition at $(\mu_B^E= 720 {\rm MeV}, T^E=93 {\rm MeV})$. The extracted freeze-out line from heavy ion collisions is close to the chiral phase transition boundary in the rPNJL model, and the kurtosis $\kappa \sigma^2$ of baryon number fluctuations from the rPNJL model along the experimental freeze-out line agrees well with the BES-I measurement. Our analysis shows that the dip structure of measured $\kappa\sigma^2$ is determined by the relationship between the freeze-out line and chiral phase transition line at low baryon density region, and the peak structure can be regarded as a clean signature for the existence of CEP. 
}
\maketitle
%
\section{Introduction}
\label{sec-1}

Exploring Quantum Chromodynamics (QCD) phase structure and understanding properties of QCD matter under extreme conditions are main topics
of heavy ion collisions. It is generally believed that there is a QCD critical end point (CEP) for chiral phase transition at finite baryon density, and searching
for the CEP and locating the CEP become one of the central goals of beam energy scan (BES) program at Relativistic Heavy Ion Collisions (RHIC),
the Facility for Antiproton and Ion Research (FAIR) in Darmstadt and Nuclotron-based Ion Collider Facility (NICA) in Dubna.

The higher order fluctuations of conserved charges carries the divergency feature of the correlation length at the CEP thus are regarded as good observables of CEP \cite{fluctuations-CEP}. The measurement on the higher cumulants of conserved charges from BES-I at RHIC for Au+Au collisions shows a non-monotonic energy dependent behavior for the kurtosis of the net proton number distributions $\kappa \sigma^2$ in the collision energy from $\sqrt{s_{NN}}=200$ to $7.7 ~{\rm GeV}$, corresponding to the baryon chemical potential $\mu_B$ from $0.1$ to $0.4$ GeV  \cite{Adamczyk:2013dal,Aggarwal:2010wy,Luo:2017faz}. It is observed that $\kappa \sigma^2$ of net proton number distributions starts from about 1 at $\sqrt{s_{NN}}=200 ~{\rm GeV}$, decreases to around 0.1 at $\sqrt{s_{NN}}=20~{\rm GeV}$ and rises quickly up to 3.5 at $\sqrt{s_{NN}}=7.7~ {\rm GeV}$. It calls for theoretical understanding whether this non-monotonic structure is related to the existence of the QCD CEP.

The QCD CEP has been investigated from lattice theory \cite{lattice}, and the recent lattice result excluded the existence of CEP in the low baryon
density region $\mu_B / T < \pi$ \cite{lattice-Frankfurt}. The QCD CEP has also been widely analyzed through symmetry analysis \cite{Symmetry-CEP},
and by using effective chiral models, e.g., the Nambu--Jona-Lasinio (NJL)  models including the Polyakov-loop improved NJL (PNJL) model \cite{NJL-CEP}, quark-meson (QM) models including the Polyakov-loop improved QM model \cite{QM-CEP}, the Dyson-Schwinger equations (DSE) \cite{DSE-CEP}, as well as the 5-dimension holographic QCD model \cite{Critelli:2017oub,Li:2017ple}. The model studies lose the power on predicting the exact location of the CEP, because different models even the same model with different parameter sets give various location of CEP. However, still we can extract some useful information from model studies, e.g., the peak structure of $\kappa \sigma^2$ along the freeze-out line can be used as a clean signature for the existence of CEP,
and the peak location of the measured $\kappa \sigma^2$ is close to the real QCD CEP mountain \cite{Li:2017ple,Li:2018ygx}.

In this talk, we show the results from a realistic Polyakov--Nambu--Jona-Lasinio (rPNJL) model \cite{Li:2018ygx}. It is interesting to observe that the kurtosis $\kappa \sigma^2$ produced from the rPNJL model along the experimental freeze-out line agrees with BES-I data well!

\section{Model setup}
\label{sec-2}

The 3-flavor rPNJL model takes into account 8-quark interaction \cite{Bhattacharyya:2016jsn}, and its effective potential is given below:
\begin{eqnarray}
\Omega & =& g_S\sum_{f}{\sigma_f^2}-\frac{g_D}{2}\sigma_u\sigma_d\sigma_s
        +3\frac{g_1}{2}(\sum_{f}{\sigma_f^2})^2+3g_2\sum_f{\sigma_f^4}-6\int_0^\Lambda \frac{d^3p}{(2\pi)^3} E_f  \nonumber \\
& & -2T\int \frac{d^3p}{(2\pi)^3} \ln[1+3(\Phi+\bar{\Phi}e^{-(E_f-\mu_f)/T})e^{-(E_f-\mu_f)/T}+e^{-3(E_f-\mu_f)/T}] \nonumber \\
& & -2T\int \frac{d^3p}{(2\pi)^3} \ln[1+3(\Phi+\bar{\Phi}e^{-(E_f+\mu_f)/T})e^{-(E_f+\mu_f)/T}+e^{-3(E_f+\mu_f)/T}] \nonumber \\
& & +U'(\Phi,\bar{\Phi},T).
\end{eqnarray}
Where $f$ takes $u,d$ for two light flavors while $s$ for strange quark, $\sigma_f=\left\langle\bar{\psi}_f\psi_f \right\rangle $ is quark-antiquark
condensate for different flavors, and $E_f=\sqrt{p^2+M_f^2}$ with $M_f$ the dynamically generated constituent quark mass taking the form of
\be
M_f=m_f-2g_S\sigma_f+\frac{g_D}{4}\sigma_{f+1}\sigma_{f+2}-2g_1\sigma_f(\sum_{f'}{\sigma_{f'}^2})-4g_2\sigma_f^3.
\ee
Here if $\sigma_f=\sigma_u$, then $\sigma_{f+1}=\sigma_d$ and $\sigma_{f+2}=\sigma_s$.
$U'$ takes the form of \cite{Ghosh:2007wy}
\be
\frac{U'}{T^4}=\frac{U}{T^4}-\kappa\ln[J(\Phi,\bar{\Phi})],
\ee
which describes the self-interaction of the Polyakov-loop $\Phi$ and $\bar{\Phi}$,  $\kappa$ is a dimensionless parameter,
\be
\frac{U}{T^4}=-\frac{b_2(T)}{2}\bar{\Phi}\Phi-\frac{b_3}{6}(\Phi^3+\bar{\Phi}^3)+\frac{b_4}{4}(\Phi\bar{\Phi})^2,
\ee
with $b_2(T)=a_0+a_1 \frac{T_0}{T}\exp(-a_2 \frac{T}{T_0})$, and
\be
J=(\frac{27}{24\pi^2})(1-6\Phi\bar{\Phi}+4(\Phi^3+\bar{\Phi}^3)-3(\Phi\bar{\Phi})^2).
\ee

\begin{table}[!ht]\centering
\begin{tabular}{c|c|c|c|c|c|c}\hline\centering
	$m_{u,d} ({\rm MeV})$ &  $m_s ({\rm MeV}) $ & $\Lambda ({\rm MeV})$ & $g_S\Lambda^{2}$  & $g_D\Lambda^{5}$  & $g_1 ({\rm MeV}^{-8})$     & $g_2 ({\rm MeV}^{-8})$  \\\hline
    	5.5         &   183.468    &    637.720 &    2.914          &      75.968       & $2.193\times 10^{-21}$    &  $-5.890\times 10^{-22}$\\\hline
\end{tabular}
\caption{Parameters for the NJL part in the rPNJL model. }
\label{parameters-NJL-part}
\end{table}

\begin{table}[!ht]\centering
	\begin{tabular}{c|c|c|c|c|c|c}\hline\centering
		$T_0$ (MeV) &  $a_0$ & $a_1$ & $a_2$  & $b_3$ & $b_4$ & $\kappa$  \\\hline
	 	  175       &  6.75  &  -9.8 & 0.26   & 0.805 & 7.555 & 0.1 \\\hline
	\end{tabular}
\caption{Parameters for the Polyakov loop part in the rPNJL model. }
\label{parameters-Polyakov-part}
\end{table}

The parameters of the NJL part shown in Table \ref{parameters-NJL-part} are fixed by vacuum properties,
and the parameters of Polyakov loop part listed in Table \ref{parameters-Polyakov-part} are fixed by global fitting of the pressure
density at zero chemical potential. With these parameters, the critical temperature and equation state
at zero chemical potential in the rPNJL model agree well with lattice data \cite{Bhattacharyya:2016jsn}.

\section{Numerical results}
\label{sec-3}

From the pressure $P=-\Omega$, the minus thermodynamical potential, one can obtain the net-baryon number fluctuations \cite{Luo:2017faz}
\be
\chi^B_n=\frac{\partial ^n[P/T^4]}{\partial [\mu_B /T]^n},
\ee
which gives the cumulants of baryon number distributions $C_n^B=VT^3\chi^B_n$. By introducing the variance $ \sigma^2=C_2^B$ and kurtosis $\kappa={C_4^B}/{(\sigma^2)^2}$, we can write the ratio of fourth and second order cumulants of net-baryon number fluctuations as
\be
 \kappa \sigma^2=\frac{C_4^B}{C_2^B}=\frac{\chi_4^B}{\chi_2^B}.
\label{Eq:ratios}
\ee
The value of $\kappa \sigma^2$ of baryon number fluctuation is 1 in the hadron resonance gas (HRG) phase and it takes the value
of $\kappa \sigma^2\simeq 0.068$ in the ideal free quark gas (FQG) limit at very high temperature \cite{Bazavov:2017dus}.

\begin{figure}
\centering
\includegraphics[width=0.5\textwidth]{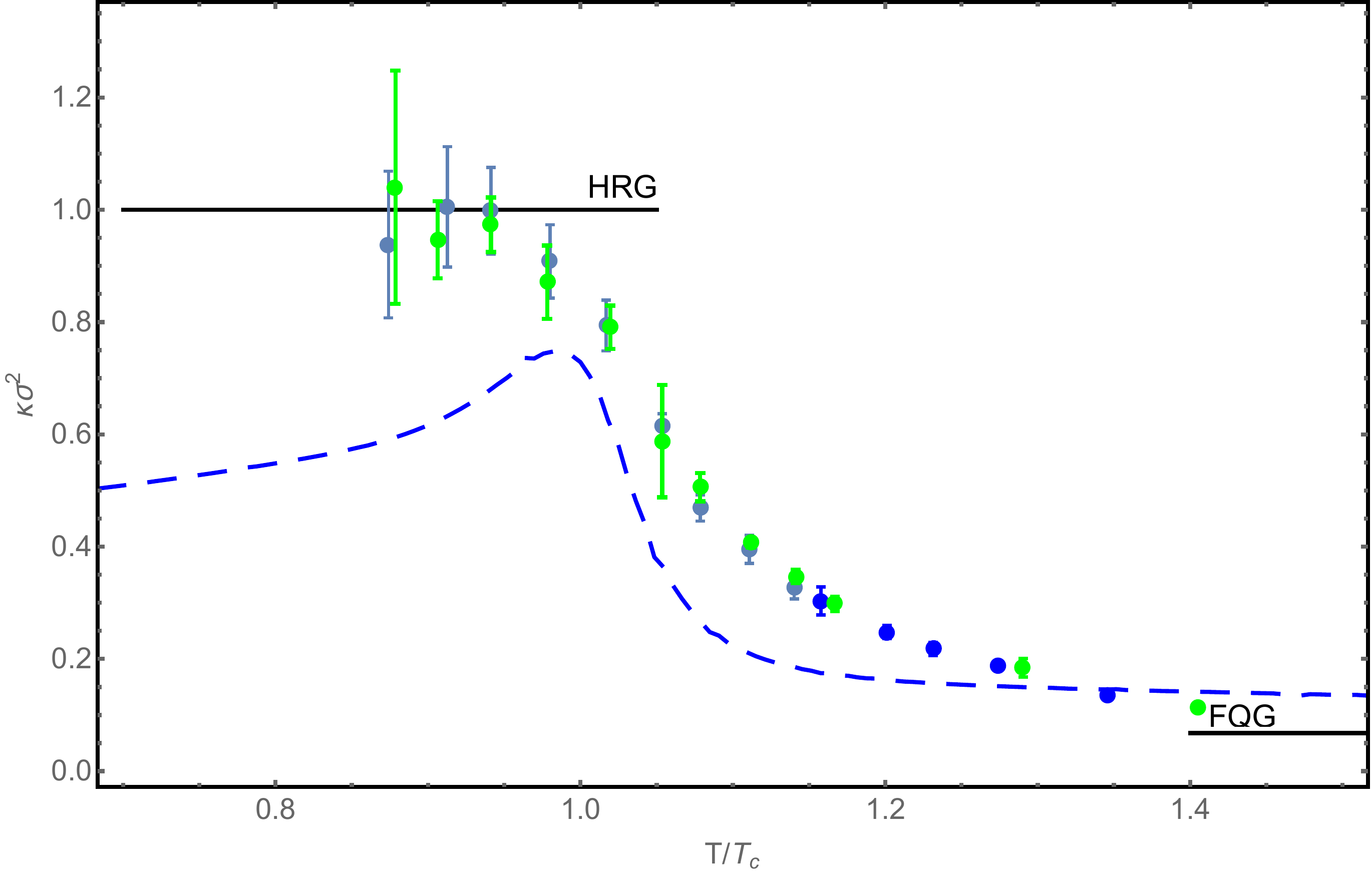}
\caption{$\kappa\sigma^2$ of baryon number fluctuation in the rPNJL model as a function of the temperature nomalized
by the chiral phase transition temperature $T_c=161 {\rm MeV}$ at zero baryon chemical potential.}
\label{fig:c4}
\end{figure}

With the parameters fixed in Table \ref{parameters-NJL-part} and  \ref{parameters-Polyakov-part}, we can obtain
$\kappa\sigma^2$ of baryon number fluctuation in the rPNJL model as shown in Fig.\ref{fig:c4} and compare with lattice data in
Ref.\cite{Bazavov:2017dus}. The HRG limit and FQG limit are also shown in Fig.\ref{fig:c4}. It is noticed that above the chiral phase transition $T_c$,
the result of $\kappa\sigma^2$ agrees well with lattice result, and below $T_c$, $\kappa\sigma^2$ of baryon number fluctuation
is only about half of the HRG limit. However, in the regular NJL model, the magnitude of $\kappa\sigma^2$ of baryon number fluctuation
is much smaller than the the lattice result \cite{Li:2018ygx}. This indicates that gluodynamics contribution plays dominant role in the baryon
number fluctuations.

\begin{figure}[h!]
\centering
\includegraphics[width=0.45\textwidth]{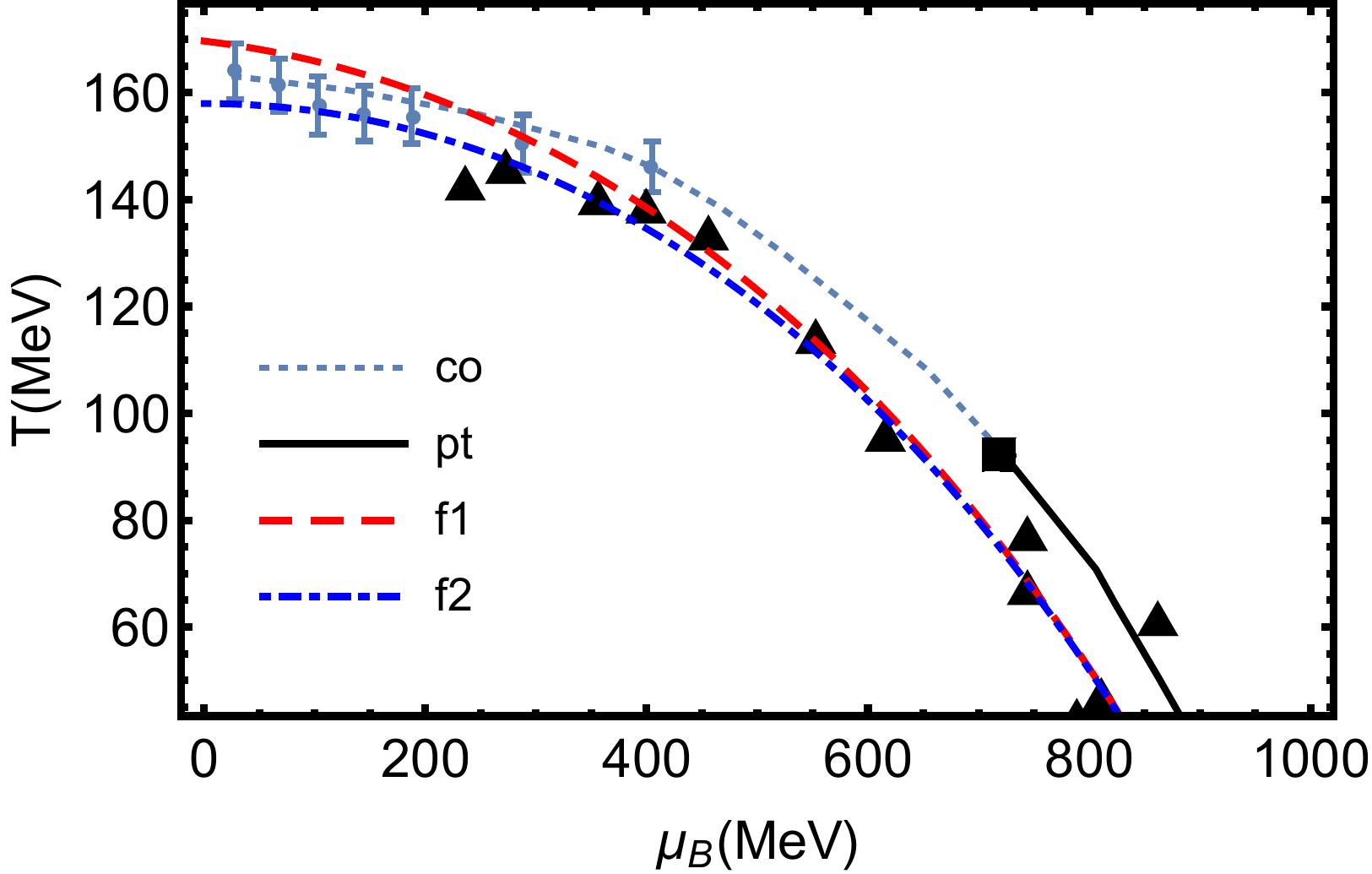}
\includegraphics[width=0.45\textwidth]{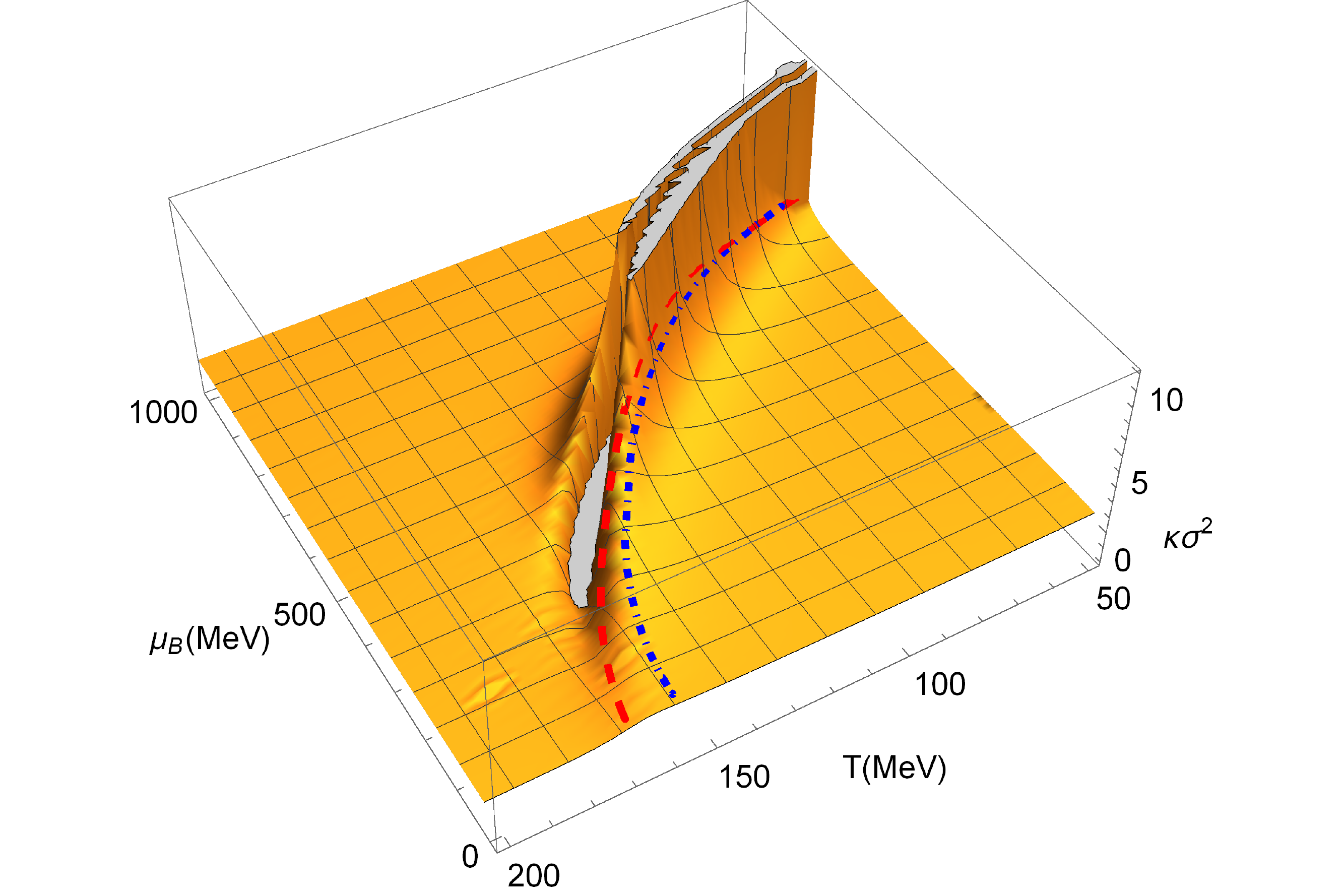}
\caption{(Left) The chiral phase boundary for the $u,d$ quarks in the rPNJL model with the CEP located at $(\mu_B^E= 720 {\rm MeV}, T^E=93 {\rm MeV})$ (marked by a square), and two freeze-out lines $f1,f2$ defined in Eq.(\ref{freezeoutline}) by fitting the freeze-out temperatures and baryon number chemical potentials extracted from BES-I at RHIC \cite{Das:2014qca} ( in dots), and for lower energy heavy-ion collisions \cite{Begun:2016pdy} (in triangles).  (Right) The 3D plot of $\kappa\sigma^2$ for the baryon number fluctuations as functions of the temperature and baryon number chemical potential in the rPNJL model, two freeze-out lines $f1,f2$ are shown in dashed and dashed-dotted lines, respectively.}
\label{fig:phasetransition-2D-3D}
\end{figure}

The chiral phase boundary in the rPNJL model is shown in Fig. \ref{fig:phasetransition-2D-3D}, with the CEP located at $(\mu_B^E= 720 {\rm MeV}, T^E=93 {\rm MeV}) $. We also show the freeze-out temperatures and baryon number chemical potentials extracted from experiment in Fig. \ref{fig:phasetransition-2D-3D}, where the freeze-out temperatures and baryon number chemical potentials extracted from BES-I at RHIC \cite{Das:2014qca} are shown in dots, and the freeze-out temperatures and baryon number chemical potentials extracted from lower energy heavy-ion collisions \cite{Begun:2016pdy} are shown in triangles. We use two fitted freeze-out lines
\bea
f1: & & T(\mu)=0.158-0.14\mu^2-0.04\mu^4-0.013(0.948-\mu)^2, \nonumber \\
f2: & & T(\mu)=0.158-0.14\mu^2-0.04\mu^4.
\label{freezeoutline}
\eea
which are shown in dashed and dashed-dotted lines, respectively. The second freeze-out line is taken from Ref.\cite{Luo:2017faz}. Both fitted experimental freeze-out lines are very close to the chiral phase boundary from the rPNJL model. The only difference is that at low baryon density region, the fitted experimental freeze-out line $f1$ starts from above the chiral phase boundary, and $f2$ starts from below the phase boundary. The 3D plot of the kurtosis $\kappa\sigma^2$ as functions of the temperature and chemical potential calculated from the rPNJL model is shown in the right of Fig. \ref{fig:phasetransition-2D-3D},
from which it can be clearly seen that the first freeze-out line $f1$ starts from the back ridge of the phase boundary and $f2$ starts from the front ridge of the phase boundary.

\begin{figure}
	\centering
\includegraphics[width=0.45\textwidth]{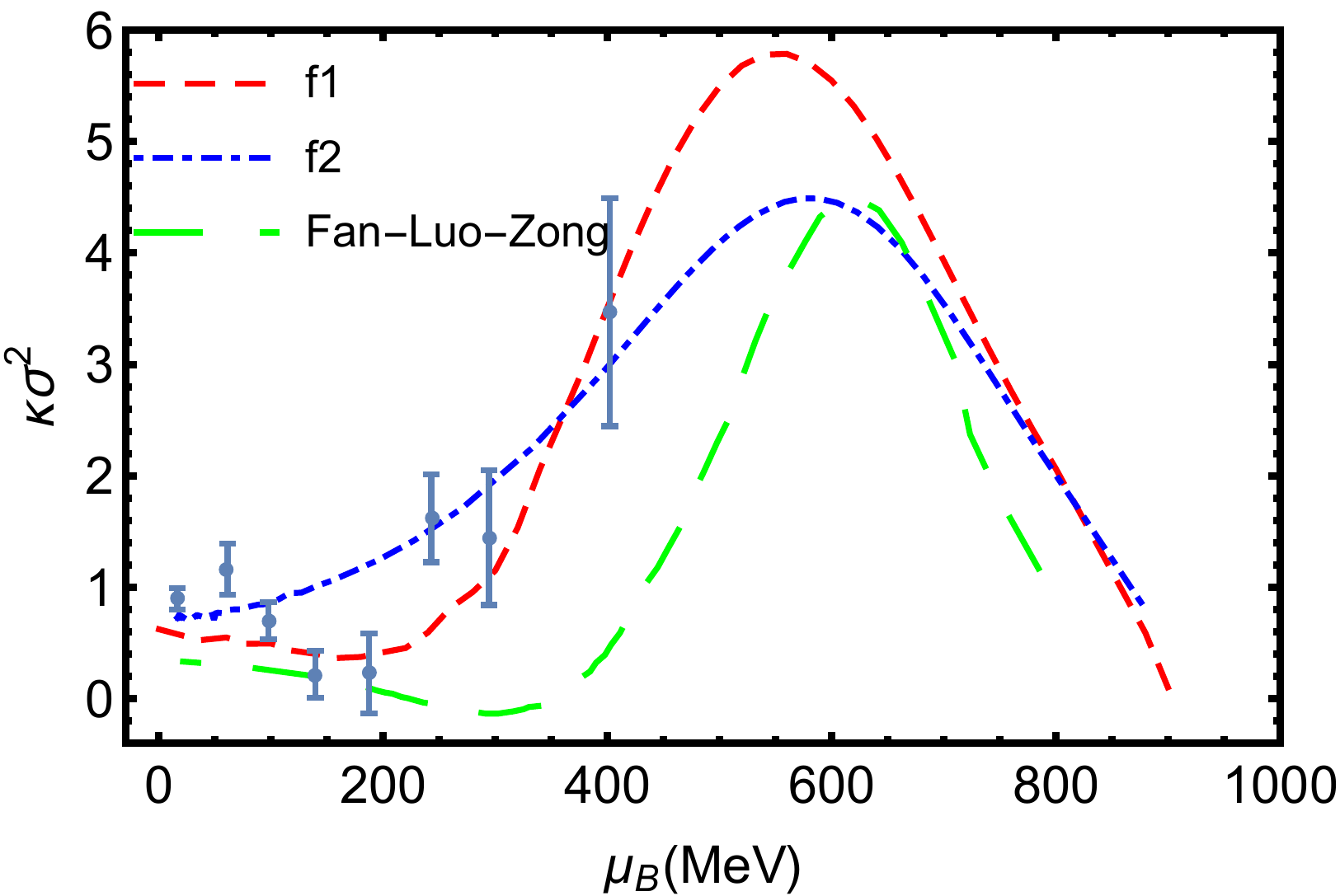}
\includegraphics[width=0.45\textwidth]{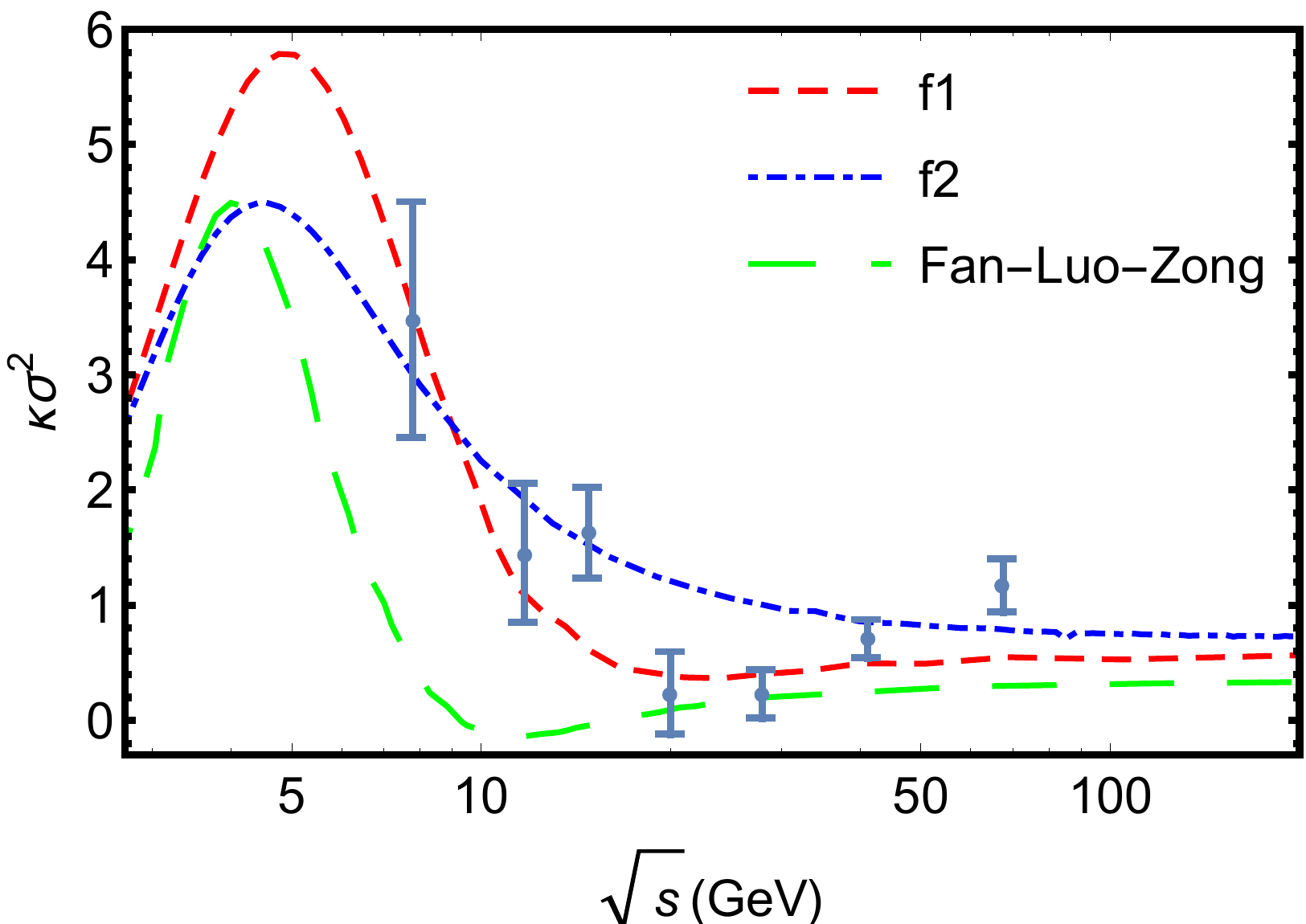}
\caption{$\kappa \sigma^2$  produced from the rPNJL model as a function of the baryon chemical potential $\mu_B$ (Left) and the collision energy $\sqrt{s}$ (Right) along the experimental freeze-out lines $f1$ (dashed line) and $f2$ (dashed-dotted line) comparing with BES-I measurement. As a reference, $\kappa\sigma^2$ from a realistic NJL model \cite{Fan:2016ovc} along the experimental freeze-out line $f2$ is shown in long dashed line. }
\label{fig:kurtosis-muB-energy}
\end{figure}

We show the kurtosis $\kappa\sigma^2$ of baryon number fluctuations calculated from the rPNJL model along the two fitted freeze-out lines
$f1,f2$ from heavy ion collisions in Fig. \ref{fig:kurtosis-muB-energy} and compare with BES-I measurement of $\kappa\sigma^2$ (shown in dots with error bars) \cite{Adamczyk:2013dal,Aggarwal:2010wy,Luo:2017faz}. The left figure is as a function of the baryon number chemical potential and the right figure is as a function of the collision energy, where the relation between the chemical potential and the collision energy \cite{Begun:2016pdy}
$\mu_{B}(\sqrt{s})=\frac{1.477}{1+0.343\sqrt{s}}$ has been used. As a reference, the $\kappa\sigma^2$ calculated from a realistic regular NJL model
along the freeze-out line $f2$ \cite{Fan:2016ovc} is also shown as long dashed line. Because there is no gluon contribution in the regular NJL model, the value of kurtosis in general is much smaller than that in the rPNJL model.

It is observed that the kurtosis $\kappa\sigma^2$ produced from the realistic PNJL model along the freeze-out line $f1$ develops a dip structure around $\mu_B=200 {\rm MeV}$ ($\sqrt{s}=20 {\rm GeV}$) and a peak structure at around $\mu_B=500 {\rm MeV}$ ($\sqrt{s}=5{\rm GeV}$, while along the freeze-out line $f2$, $\kappa\sigma^2$ only develops a peak structure at around $\mu_B=500 {\rm MeV}$ ($\sqrt{s}=5{\rm GeV}$ and no dip structure is developed. As we have discussed that the only difference between the two freeze-out lines is at the low baryon density region: $f1$ crosses the chiral phase boundary from above, i.e. starts from the back ridge of the phase boundary, from the 3D plot in Fig. \ref{fig:phasetransition-2D-3D}, it is natural to see how the dip structure is formed. For the case along the freeze-out line $f2$, which always lies below the phase boundary, there is no chance to form the dip structure. The peak structure along the freeze-out line is solely determined by the existence of the CEP. From this analysis, we can see that the dip structure of measured $\kappa\sigma^2$ is determined by the relationship between the freeze-out line and chiral phase transition line at low baryon density region, and the peak structure can be regarded as a clean signature for the existence of CEP.

Another unexpected result is that the kurtosis $\kappa\sigma^2$ from the equilibrium result in the rPNJL model along the experimental freeze-out line $f1$ agrees with BES-I measurement very well. This is a surprising result, because in general, one should take into account the non-equilibrium evolution of the system in heavy ion collisions. However, as shown in Ref.\cite{Memory-Yin} that the kurtosis of baryon number fluctuation even changes sign and
becomes negative. It is worthy of mentioning that the measurement of $\kappa\sigma^2$ is always positive, therefore it deserves further understanding on how the "little bang" after collision evolves with time. As shown in \cite{BraunMunzinger:2003zd}, one possible scenario is that the system after collision reaches thermalization very quickly in quite high temperature and then evolves in equilibrium state. From our analysis, to form the dip structure, the freeze-out temperature at low baryon density region should be higher than the phase transition temperature! If this scenario is correct, the equilibrium result from a realistic PNJL model might work. 

\section{Summary}
\label{sec-4}
In summary, we have investigated the kurtosis of the baryon number fluctuations $\kappa \sigma^2$ in a reparameterized realistic PNJL model, in which the critical temperature, equation of state and baryon number fluctuations are in good agreement with lattice data at zero chemical potential. This rPNJL model predicts a CEP located at $(\mu_B^E= 720 {\rm MeV}, T^E=93 {\rm MeV})$. The extracted experimental freeze-out line from heavy ion collisions is very close to the chiral phase transition boundary in the rPNJL model.  It is surprised to see that the kurtosis $\kappa \sigma^2$ of baryon number fluctuations from the rPNJL model along the experimental freeze-out line agrees well with the BES-I measurement. We also analyzed the formation of the dip structure and peak structure of measured $\kappa\sigma^2$ along the freeze-out line. Our analysis shows that the dip structure of $\kappa\sigma^2$ is determined by the relationship between the freeze-out line and chiral phase transition line at low baryon density region, and the peak structure is solely determined by the divergency at CEP, and the peak structure can be regarded as a clean signature for the existence of CEP. It is worthy of mentioning that at low baryon density region, the extracted freeze-out temperatures from BES-I measurement at RHIC \cite{Das:2014qca} are indeed higher than the critical temperatures, this supports our analysis on the formation of dip structure of $\kappa \sigma^2$ along the freeze-out line.

Our result shows that the equilibrium result can explain the BES-I data on baryon number fluctuations, this may indicate that the system after collision reaches thermalization quickly and evolve in equilibrium before the freeze-out and phase transition, which is in agreement with the analysis in \cite{BraunMunzinger:2003zd}. For example, in the collision energy of $\sqrt{s}=200 {\rm GeV}$, the system reaches thermalization at around the temperature of $T\simeq 210-230 {\rm MeV}$ , which is much higher than the freeze-out temperature and the phase transition temperature around $160 {\rm MeV}$. 

\section*{Acknowledgments}
\label{acknowledgments}
This work was supported in part by the NSFC under Grant Nos. 11725523, 11735007, and 11261130311 (CRC 110 by DFG and NSFC).

\end{document}